\documentclass[MSNbibl,nameyear,dvips]{arxstspdf}
\usepackage{flushend}
\usepackage{stfloats}
\usepackage{graphicx}

\volume{27}
\issue{3}
\pubyear{2012}
\firstpage{340}
\lastpage{343}
\doi{10.1214/12-STS381A} 
\referstodoi{10.1214/11-STS381}
\setattribute{copyright}{owner}{In the Public Domain}

\begin{document}
\begin{frontmatter}
\vspace*{6pt}
\title{Discussion of ``Multivariate
Bayesian Logistic Regression for Analysis of Clinical Trial Safety
Issues'' by W.~DuMouchel}
\runtitle{Comments on MBLR}

\begin{aug}
\author[a]{\fnms{Bradley W.} \snm{McEvoy}\ead[label=e1]{Bradley.McEvoy@fda.hhs.gov}}
\and
\author[b]{\fnms{Ram C.} \snm{Tiwari}\corref{}\ead[label=e2]{Ram.Tiwari@fda.hhs.gov}}
\runauthor{B. W. McEvoy and R. C. Tiwari}

\affiliation{Center for Drug Evaluation and Research}

\address[a]{Bradley McEvoy is a
reviewer in Division of Biometrics VII, Office of Biostatistics, Center
for Drug Evaluation and Research, FDA, 10903 New Hampshire Avenue,
Silver Spring, Maryland 20993-0002, USA \printead{e1}.}
\address[b]{Ram Tiwari is
Associate Director in the Office of Biostatistics, Center for Drug
Evaluation and Research, FDA, 10903 New Hampshire Avenue, Silver Spring,
Maryland 20993-0002, USA \printead{e2}.}

\thankstext[]{ss}{The views expressed by authors are their own and do not
necessarily reflect those of FDA.}

\end{aug}



\end{frontmatter}

We would like to comment on this article by Wil-liam DuMouchel, as it
gives an interesting application of logistic regression to clinical
safety data. Not to underscore the scope of the multivariate Bayesian
logistic regression (MBLR) model, but the use of numerical integration
is arguably its most important feature. Avoiding Markov chain Monte
Carlo (MCMC) sampling techniques for other data-mining tools, such as
the Multiple-item Gamma Poisson Shrinker (DuMouchel, \citeyear{r2}), has proven
successful for Dr. DuMouchel in their acceptance among nonstatisticians.
With MBLR this should not be an exception.

As most statisticians lack the clinical insight required to specify the
appropriate MBLR model inputs, it makes MBLR an ideal tool for use by
the clinicians. However, targeted users may not appreciate some
subtleties of MBLR, which we present below. We also present findings
from our empirical evaluation of the MBLR algorithm. This commentary
provides some perspective that we have gained through multiple
interactions with Dr. DuMouchel and from our reviews of different
versions of MBLR formulation at FDA since 2009.

\section{MBLR and Meta-Analysis}
In order to fully appreciate the MBLR methodology, one has to contrast
it with a more traditional meta-analytical formulation when data from
multiple trials are investigated. Dr. DuMouchel is correct in pointing
out that the MBLR methodology is in the spirit of a full-data
meta-analysis and does not consider it a meta-analytic model. The
current MBLR model formulation does not render the flexibility of
separating out patient- and trial-level variations in the model.
Consequently, MBLR is very different from a multi-level/meta-analysis
model that would consist of a patient-level model and a trial-level
model, each with independent sources of variation. This makes MBLR
effectively a patient-level model; the inclusion of trial-level
variables (e.g., study identifiers) into equation (2) results in the
variance components in equations (3)--(6) being influenced by both
patient and trial heterogeneity.

This distinction between the MBLR and its meta-analytic formulation is
critically important. The\break main advantage of a meta-analytic formulation
is that it preserves the trial-specific randomized comparison between
the treatment and control groups, thereby avoiding confounded estimates.
With the MBLR formulation this is not necessarily the case, as Dr.
DuMouchel aptly notes for the Pollakiuria example that the
trial-specific estimates do not preserve the between-trial differences.
Additionally,\break shrinkage estimates used to identify vulnerable patient
subgroups depend on factors which are typically considered unrelated of
patient characteristics.

The practical concern of applying a methodology that does not ensure the
randomized comparison is preserved is that it may lead to a possible
signal being missed or hidden. A recent high-profile example of this
concern was the meta-analysis of the diabetes drug rosiglitazone (Rucker
and Schumacher, \citeyear{r5}). When safety data collected from the randomized
controlled trials were pooled by trial arm, it resulted in Simpson's
paradox.

It is, therefore, important to understand the subtle distinction of how
MBLR differs from the more traditional meta-analytic models, and the
potential consequences that may arise from the use of MBLR.
Unfortunately, the MBLR tool/program in its current capacity does not
have the capability to evaluate the \mbox{potential} implications discussed in
the aforementioned paragraphs. This necessitates the use of other
statistical methodologies to fully evaluate the results from MBLR
software, which, paradoxically, is the situation that Dr. DuMouchel
initially set out to avoid. That said, it would be a nice extension if
the MBLR methodology was expanded, incorporating the suggestions
outlined above, thereby increasing the general utility of the tool.
Next, we present an attempt toward this extension.
\begin{figure*}

\includegraphics{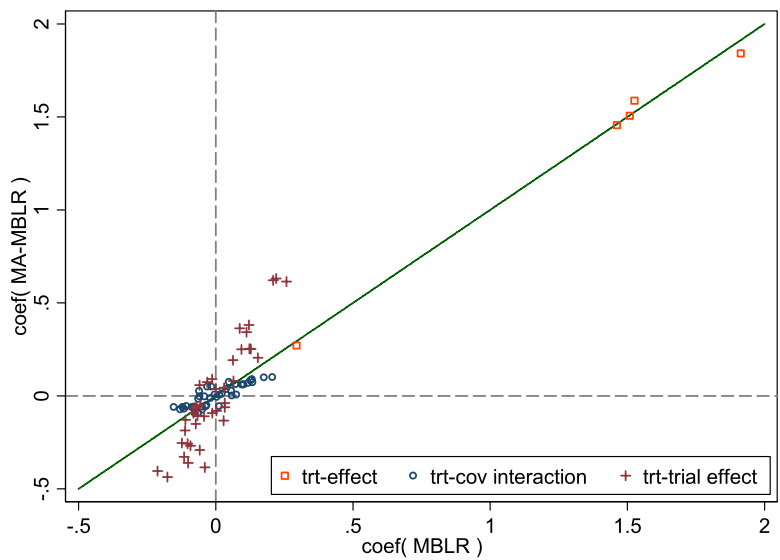}

  \caption{Relationship between MA-MBLR and MBLR for selected model
parameters.}\label{fig1}\vspace*{-3pt}
\end{figure*}

\section{Meta-Analytic MBLR Formulation}

We present a modified MBLR model motivated from a meta-analytic
perspective, which we shall, henceforth, refer to as meta-analytic MBLR
(MA-MBLR). Using the notation from the paper, let the~$J$ covariates
correspond only to patient-level characteristics and assume that there
are a total of $L$ trials. Then, the MA-MBLR patient-level model for trial
$l$, $l=1,\ldots,L$, and issue $k$ is given by
\begin{eqnarray*}
\operatorname{logit}(p_{ikl}) &=& \alpha_{0kl} + \sum_{g}
X_{igl}\alpha_{gk}
\\
&&{} + T_{il} \biggl(\beta_{0kl} +
\sum_{g} X_{igl}\beta_{gk}\biggr).
\end{eqnarray*}
Unlike the MBLR formulation, the MA-MBLR would assume the trial-specific
intercept $\alpha_{0kl}$ and treatment effect $\beta_{0kl}$ have
distinct variance components, thereby separating patient and trial
variability. This can be formally achieved by assigning the
trial-speci\-fic~intercept and treatment effect of the following
hierarchical prior: $\alpha_{0kl}\! \sim\! N(\alpha_{0k} ,
\sigma^{2}_{A.k})$ and $\beta_{0kl} \!\sim\! N(\beta_{0k}
,\break \sigma^{2}_{0.k})$, for $k=1,\ldots,K$ and $l=1,\ldots,L$. The MA-MBLR
model is fully specified by
equations
(3)--(6), as well as by the hyperpriors for the model's hyperparameters, and
has the ($2K+4$) standard de\-viations, ($\sigma_{A.1}, \ldots,
\sigma_{A.K}, \sigma_{0.1},\allowbreak   \ldots, \sigma_{0.K}, \sigma_{A},
\sigma_{0},\sigma_{B}, \tau)$,\break that have independent
uniform distribution on the interval 0 to $d$, as specified in the
paper.\vadjust{\goodbreak}

We investigated for the data-example in the paper whether the MBLR and
MA-MBLR formulations make a substantive impact on the risk assessment
for the five most frequent issues. Both the MBLR and MA-MBLR models were
fit using OpenBUGS (Lunn et al., \citeyear{r3}), and thus are fully Bayesian MBLR
and MA-MBLR. The fully Bayesian models differed from the MBLR model
described in the paper in three ways, namely, (i) it assumes diffused
normal priors for the location parameters rather than uniform
noninformative priors, (ii) it constrains the hyperpriors $A_{g}$ such
that the $g_{j}$th level of covariate $j$ is equal to the negative sum
of the remaining $g_{j}-1$ levels, and (iii) the support of the prior
for the standard deviation $d$ was increased to 3.

Figure \ref{fig1} shows the relationship for some of the estimated parameters.
The issue specific treatment effect $\beta_{0k}$ did not differ too
much between models. However, the interaction term between treatment and
the patient-level covariates tended to be closer to the null value for
MA-MBLR, while the MA-MBLR trial-specific treatment effect tended to be
further away from the null value than MBLR. Although there were no
surprising differences noted between the MBLR and MA-MBLR coefficients
for this example, the two different formulations can possibly result in
different substantive conclusions.\vspace*{-3pt}

\begin{figure*}[b]
\vspace*{-3pt}
\includegraphics{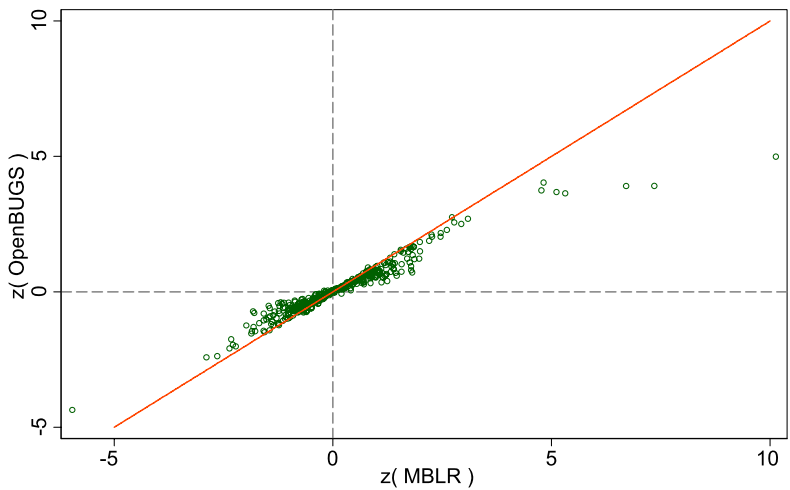}

  \caption{Relationship of $z$-scores from fully Bayesian model fit using
OpenBUGS compared to MBLR.}\label{fig2}
\end{figure*}

\section{Borrowing Information Across Issues}\vspace*{-3pt}

It is important to note that MBLR borrows information across issues by
positing a hierarchical distribution to parameters from parallel
logistic regression models, and \textit{does not} model the joint
distribution of the endpoints. An example of the latter approach is
given by Bayesian multivariate logistic regression (O'Brien and Dunson,
\citeyear{r4}). More importantly, there needs to be recognition among its users
that an analysis that borrows information across issues is not
inherently better than the one that does not.

To illustrate a possible peril of borrowing information across issues,
suppose the issues selected are medically related, but they vary in
their severity; in particular, assume there is one severe issue that
occurs infrequently and the remaining issues are less severe but occur
more frequently. Because the amount of information borrowed across
issues from MBLR is related to the precision of the estimate (which is a
function of the issue frequency), the effect for the less frequent
issues would be sensitive to the effects for the more frequent issues.
It is important that users of the tool are mindful of such
considerations.\vadjust{\goodbreak}

\section{MBLR Estimation Algorithm}\vspace*{3pt}
As stated previously, we believe the advantage of the MBLR methodology
is in obtaining posterior inferences that do not rely on computationally
time-consuming estimation methods (such as MCMC methods). However, the
timeliness of the analysis has to be balanced by the well-known
limitations of the Laplace approximation of the integral of the
posterior density (Carlin and Louis, \citeyear{r1}), which are applicable to
MBLR.

As part of the software review at FDA, we evaluated the adequacy of
MBLR's estimation algorithm by contrasting results obtained from the
fully Bayes\-ian MBLR using OpenBUGS; the comparison was based on the data
described in the paper. The fully Bayesian MBLR differed from the MBLR
by points~(i) and (ii) listed above. The two estimation approaches
yielded similar esti\-mates for the variance components $\varphi =
(\sigma_{A},\sigma_{0}, \sigma_{B} ,\tau)$ and the
parameter estimates had almost perfect correlation ($\rho = 0.9998$).
However, the relationship based on $z$-scores ($=$estimate$/$\break standard error),
presented in Figure \ref{fig2}, suggests that MBLR has smaller standard errors
than the full Bayesian analysis. This observation is also supported by
the simulation results, where MBLR tended to have a type-I error rate
that slightly exceeded the nominal 10\%
level.\vadjust{\goodbreak}

\section{Conclusion}\vspace*{-2pt}

The MBLR model will have a profound impact as it is rolled-out being
used for clinical safety data analysis. However, in order to realize
MBLR's potential strengths and pitfalls, it will require collaboration
between its different user-constituents, those being statisticians and
subject-matter experts.\vspace*{-2pt}


\end{document}